\documentclass[conference]{IEEEtran}
\usepackage{graphicx}
\usepackage{amsmath}
\usepackage{multicol}
\usepackage{multirow}
\usepackage{stfloats}
\usepackage{booktabs}
\usepackage{mathrsfs}
\usepackage{enumerate}
\usepackage{amssymb}
\usepackage{algorithm}
\usepackage{array}
\usepackage{underscore}
\usepackage{url}
\usepackage{hyperref}
\hypersetup{hidelinks}
\usepackage{flushend}

\usepackage{algorithmic}
\usepackage[square, comma, sort&compress, numbers]{natbib}
\usepackage{graphicx}
\usepackage{epstopdf}
\usepackage{caption}
\usepackage{subcaption}

\usepackage{color}

\usepackage[outerbars,color]{changebar}
\ifx\pdfoutput\undefined
\else\ifnum\pdfoutput>0
  \usepackage{pdfcolmk}
\fi\fi
\cbcolor{red}
\usepackage{fancyhdr}

\makeatletter
\newcommand{\Rmnum}[1]{\expandafter\@slowromancap\romannumeral #1@}
\makeatother

\usepackage{array}  \newcommand{\PreserveBackslash}[1]{\let\temp=\\#1\let\\=\temp}  \newcolumntype{C}[1]{>{\PreserveBackslash\centering}p{#1}}  \newcolumntype{R}[1]{>{\PreserveBackslash\raggedleft}p{#1}}  \newcolumntype{L}[1]{>{\PreserveBackslash\raggedright}p{#1}}

\ifCLASSINFOpdf
\else
\fi
\begin{document}
\title{Adaptive Distributed Laser Charging \\for Efficient Wireless Power Transfer}

\author{\IEEEauthorblockN{Qingqing~Zhang\IEEEauthorrefmark{1},
Xiaojun Shi\IEEEauthorrefmark{2},
Qingwen~Liu\IEEEauthorrefmark{1},
Jun~Wu\IEEEauthorrefmark{1},
~Pengfei~Xia\IEEEauthorrefmark{1} and
Yong Liao\IEEEauthorrefmark{3}}
\IEEEauthorblockA{\IEEEauthorrefmark{1}Dept. of Computer Science and Technology, Tongji University, Shanghai, China,\\
\IEEEauthorrefmark{2}China Electronics Technology Group Corporation, Beijing, China,\\
\IEEEauthorrefmark{3}China Academy of Electronics and Information Technology, Beijing, China.}
\IEEEauthorblockA{Email: zhangqingqing033@gmail.com, worldjun31@163.com, qingwen.liu@gmail.com,\\
wujun@tongji.edu.cn, pengfei.xia@gmail.com, yliao@csdslab.net}}

\IEEEspecialpapernotice{\vspace{0.0em}}
\maketitle

\begin{abstract}
Distributed laser charging (DLC) is a wireless power transfer technology for mobile electronics. Similar to traditional wireless charging systems, the DLC system can only provide constant power to charge a battery. However, Li-ion battery needs dynamic input current and voltage, thus power, in order to optimize battery charging performance. Therefore, neither power transmission efficiency nor battery charging performance can be optimized by the DLC system. We at first propose an adaptive DLC (ADLC) system to optimize wireless power transfer efficiency and battery charging performance. Then, we analyze ADLC's power conversion to depict the adaptation mechanism. Finally, we evaluate the ADLC's power conversion performance by simulation, which illustrates its efficiency improvement by saving at least 60.4\% of energy, comparing with the fixed-power charging system.

\end{abstract}

\IEEEpeerreviewmaketitle
\section{Introduction}\label{Section1}
Power consumption of mobile devices is increasing rapidly due to the fast-growing demand of multimedia computation and communication. Hence, battery has occupied more than 70\% weight and space of current smartphones. Moreover, people are suffering the headache of always carrying a power cord and seeking outlet. In this case, wireless charging becomes an attractive solution for these problems.

Nikola Tesla's pioneering electricity transmission over the air was presented in 1904. Since then, many wireless charging or wireless power transfer technologies have been proposed including magnetic inductive coupling, magnetic resonance coupling, microwave radiation, heat radiation, laser, ultrasound, etc. However, safely transmitting Watt-level power over meter-level distance for mobile electronics is still challenging \cite{liu2016dlc}.

\begin{figure}
	\centering
	\includegraphics[width=3.1in]{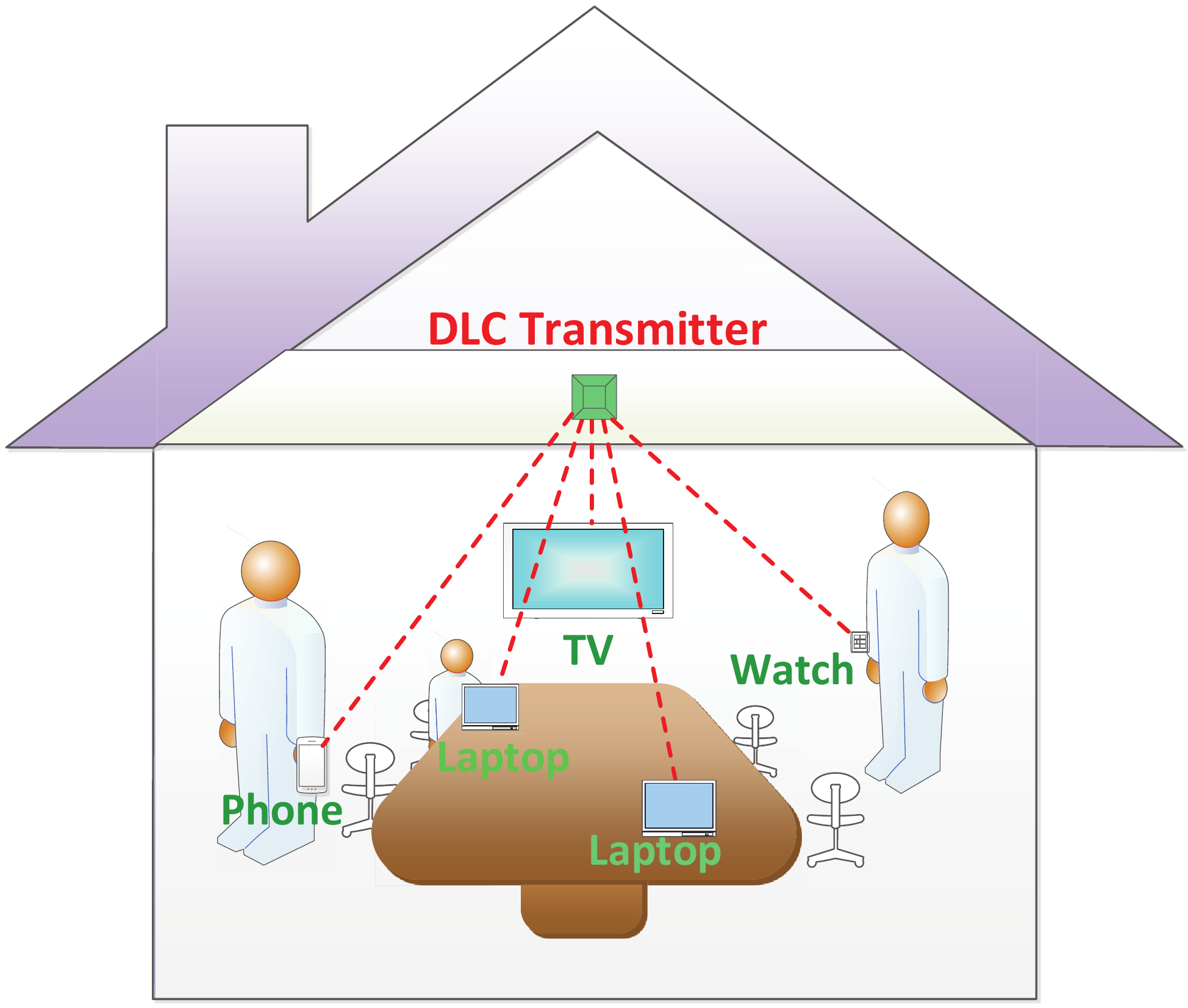}
	\caption{DLC Application Example}
	\label{infrastructural}
\end{figure}

Recently, an infra-distributed-laser based wireless charging technology was innovated \cite{wi-charge}. It can safely transfer 2W-power over 5m-distance for multiple mobile devices. Such DLC technology supports mobile power transfer like WiFi as in Fig.~\ref{infrastructural} \cite{liu2016dlc}. The DLC's mechanism, feature, application, etc. have been presented in \cite{liu2016dlc}, which outlines the brilliant future of mobile charging for electronic devices anywhere and anytime.

Li-ion battery is the widely-used battery for mobile devices. It needs dynamic input current and voltage for battery performance optimization, e.g., battery capacity maximization \cite{charging lithium ion}. However, the DLC system can only transmit predetermined and fixed power \cite{liu2016dlc}. Therefore, DLC's power transfer efficiency and battery charging performance are not optimized. In this paper, we present an ADLC system design, which can optimize the power transmission efficiency and Li-ion battery charging performance.

The contribution of this paper includes: 1) we propose the ADLC system, which can automatically adjust laser power relying on feedback control mechanism; 2) we adopt a direct current to direct current (DC-DC) converter in the ADLC system to deliver the desired voltage and current for battery charging; 3) we analyze power conversion in the ADLC system, which is designed to optimize laser power utilization and battery charging performance.

\section{ADLC Architecture}\label{Section2}
In this section, we will briefly describe the DLC system. Then, we will present the detailed ADLC system design.

\subsection{DLC System}\label{}
\begin{figure}
	\centering
	\includegraphics[width=3.25in]{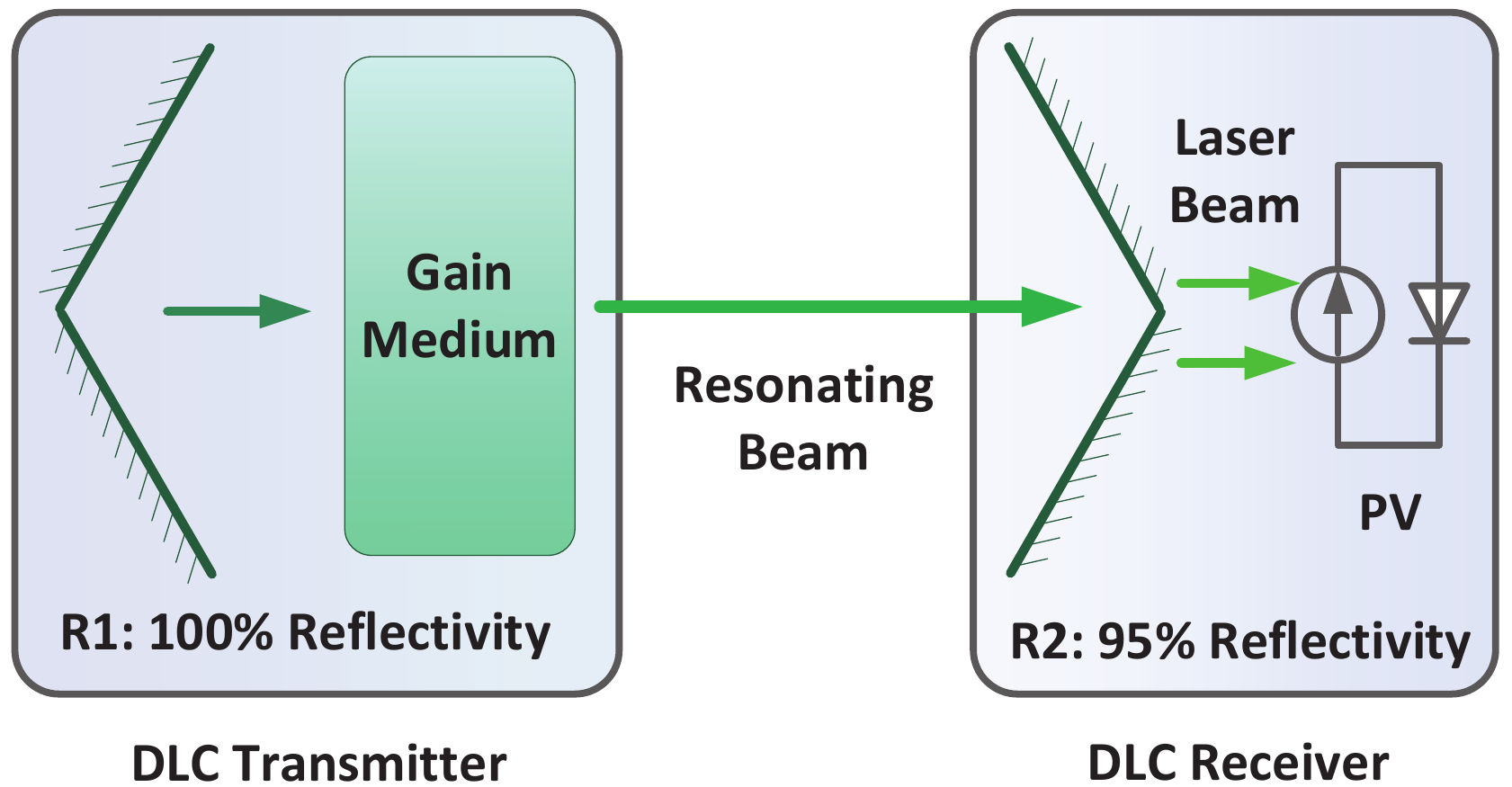}
	\caption{Distributed Laser Charging System}
	\label{distributedresonatinglaser}
\end{figure}

Traditional laser is generated by a resonator that consist of two parallel mirrors and a gain medium between the two mirrors, which are integrated into a single device \cite{photonic}. However, in the DLC system, these resonator components are separated as shown in Fig. \ref{distributedresonatinglaser}.

Fig. \ref{distributedresonatinglaser} shows the DLC system diagram as described in \cite{liu2016dlc}. In the DLC transmitter, there is a retro-reflector mirror R1 with 100\% reflectivity and a gain medium to amplify passing-by photons. Whereas  in the DLC receiver, a retro-reflector mirror R2 with 95\% reflectivity is contained. Photons that pass through R2 can form the laser beam. A photovoltaic panel (PV-panel) is installed behind the mirror R2. The laser beam can be converted to electricity by the PV-panel, which is similar to a solar panel. Due to the distributed resonator structure, the name of ``distributed resonating laser'' is adopted in DLC systems. DLC is the wireless charging technology based on distributed resonating laser.

It is well-known that laser can convey high-power energy. However, safety and alignment are the major concerns in traditional laser power transfer for mobile electronics. In DLC systems, any object blocking the line of sight (LOS) between R1 and R2 can break resonation, which provides the inherent safety. In addition, as long as photons travel along the LOS connecting R1 and R2, they can establish resonation without concerning about the incident angle. Hence, resonating beam can be self-aligned without specific aim or track. These two DLC features overcome the difficulties of traditional laser, which enables mobile charging safely.

\subsection{ADLC System}\label{}
Li-ion battery is the most widely-used rechargeable battery for mobile devices. For Li-ion battery, even slightly undercharging can lead to significant capacity reduction. On the other hand, overcharging may damage the battery or even cause danger \cite{Dearborn2005}. In this case, offering controllable voltage and current is important to charge Li-ion battery safely and achieve its optimal performance. ``Battery charge profile'' refers to changing characteristics of voltage-current set values during the battery charging period, it aims to obtain the optimally charging performance. The DLC system only provides fixed and predetermined laser power, which cannot meet the dynamic battery charge profile.

Furthermore, in the DLC receiver, if the output power from PV-panel is not fully stored into battery, the extra energy usually causes thermal effect. This leads to efficiency reduction, battery damage, or even danger.

In order to deal with these issues, an intuitive idea is to adaptively transmit power at the DLC transmitter based on the feedback information from the DLC receiver. The similar mechanism for signal transmission, namely link adaptation, is well-known in wireless communications, in order to optimize the information delivery.
\begin{figure}
	\centering
	\includegraphics[width=3.4in]{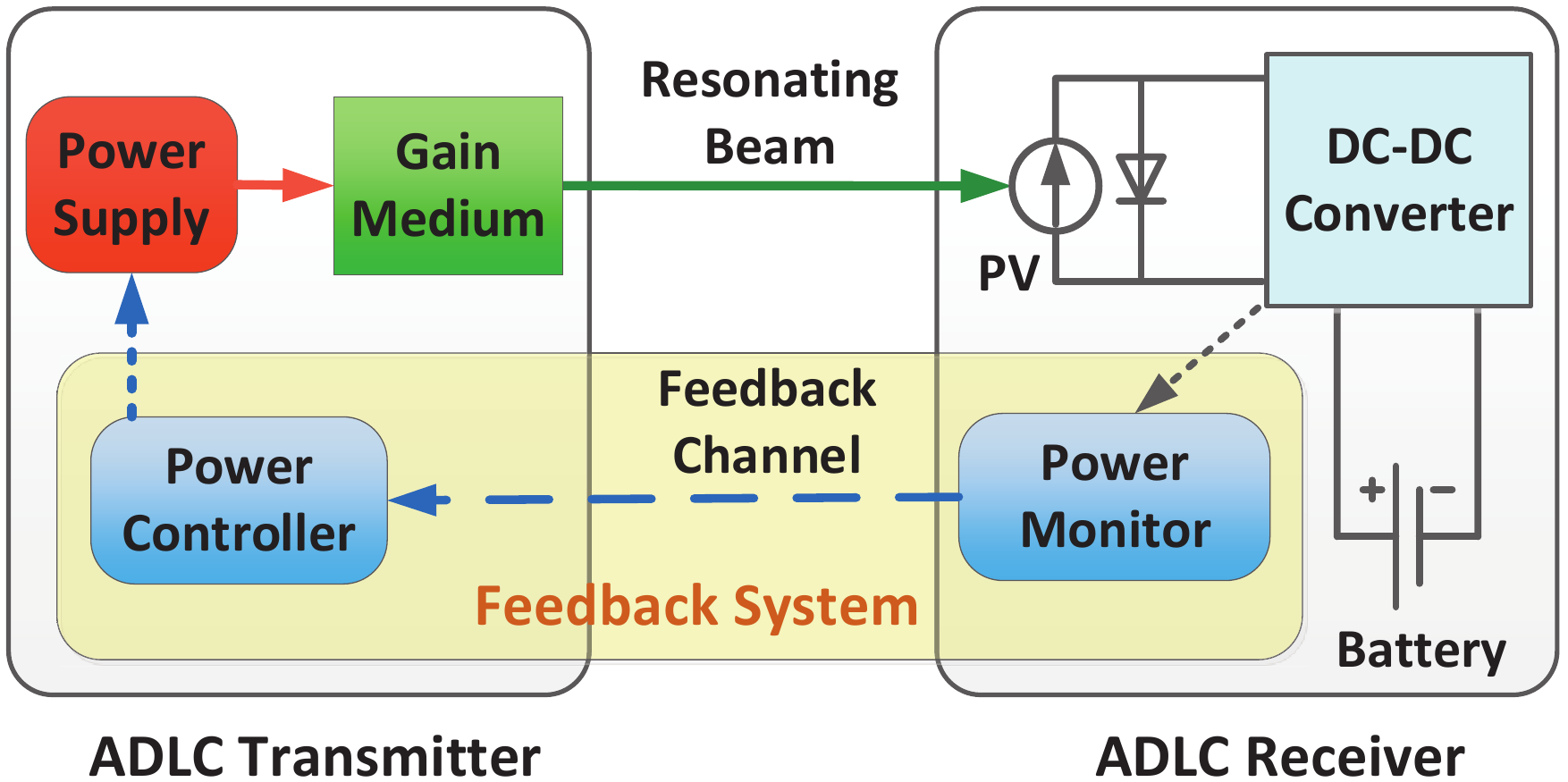}
	\caption{Adaptive Distributed Laser Charging System}
	\label{adaptivepower}
\end{figure}

In Fig. \ref{adaptivepower}, we present here the ADLC power transmission system diagram. The modules in the ADLC system include: power supply and gain medium at the transmitter; PV-panel, DC-DC converter, and battery at the receiver. The feedback channel is across the two ends, which consists of power monitor and power controller.

\section{Adaptation Mechanism}\label{Section3}
In this section, we will at first introduce the charge profile of Li-ion battery. Based on the charge profile and the battery status tracked by the power monitor, the desired laser power can be derived. Then, this desired laser power can be sent to the power controller through the feedback channel, which will indicate the power supply to stimulate the corresponding laser power. At the receiver, laser could be converted to electricity by PV-panel. Then the DC-DC converter will change PV-panel electrical output to the desired voltage-current values to charge the battery optimally.

For analytical simplicity, we take the following assumptions: \\
1) Power loss due to resonating beam propagation over the air can be ignored, which could be validated in clear air and meter-level distance \cite{Isaac2009}. \\
2) Laser diameter is constant, so that laser power depends on laser irradiation only. This could be validated by controlling DLC transmitter and receiver aperture diameters \cite{photonic}. \\
3) DC-DC converter efficiency is 100\%, which could be validated since the typical DC-DC converter efficiency is above 90\%-95\% \cite{DC-DC converter2}.\\

\subsection{Li-ion Battery Charge Profile}\label{}
\begin{figure}
	\centering
	\includegraphics[width=3.4in]{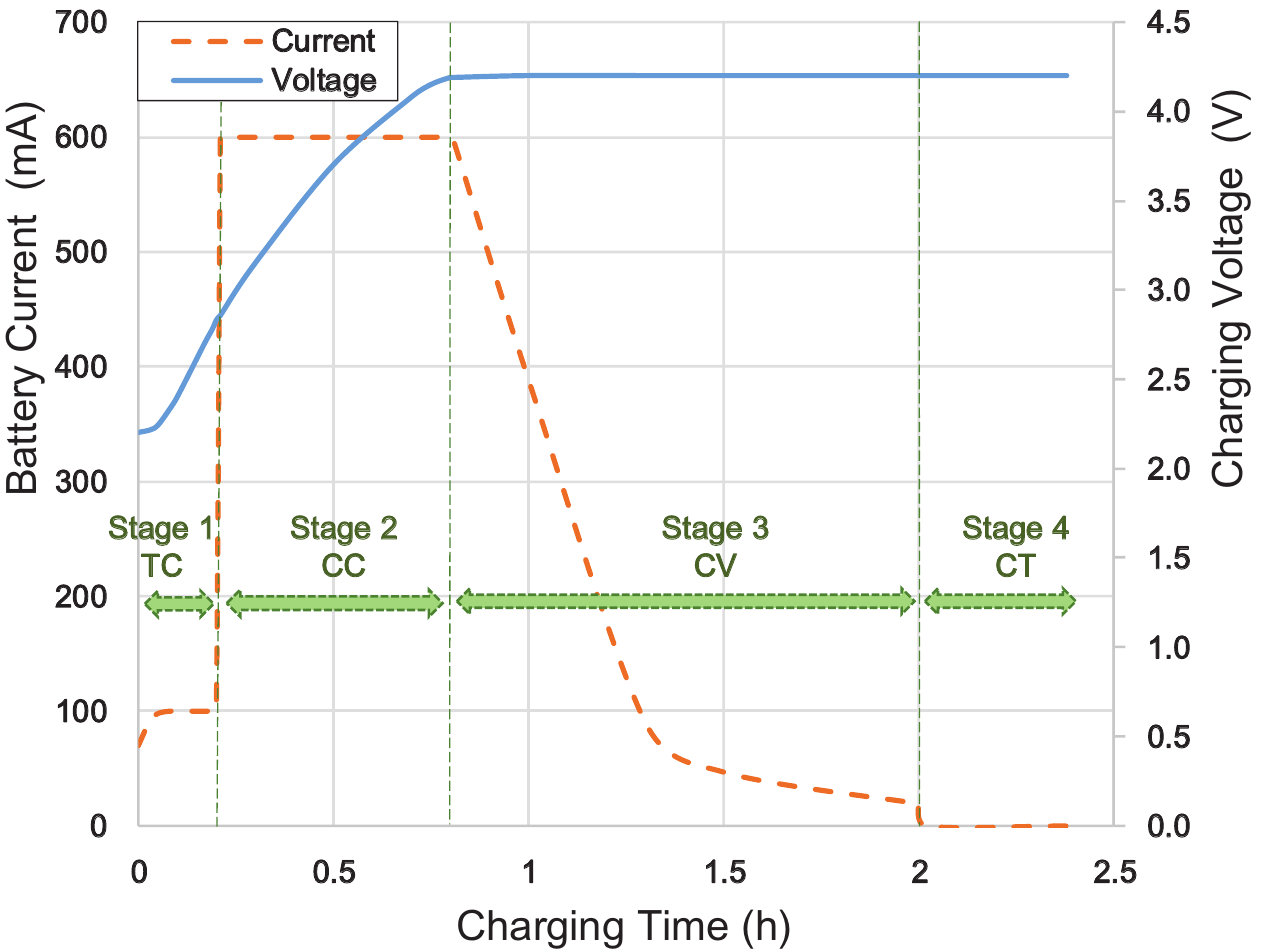}
	\caption{Li-ion Battery Charge Profile}
	\label{li-ionchargezero}
\end{figure}

Different kinds of batteries may have different charge profiles given their chemical characteristics \cite{Cleveland2008}. For example, we will discuss here the well-known Li-ion battery charge profile.
It includes four stages as in Fig. \ref{li-ionchargezero}:

Stage 1: Trickle Charge (TC) - When the battery voltage is below 3.0V, the battery is charged with an increasing trend of current towards 100mA. The voltage will raise to 3.0V.

Stage 2: Constant Current (CC) charge - After the battery voltage has risen above 3.0V, the TC-CC threshold, the charge current switches to constant value, which should be between 200mA to 1000mA. The voltage will raise towards 4.2V.

Stage 3: Constant Voltage (CV) charge - When the voltage reaches 4.2V, the CC-CV threshold, CC charge ends and CV stage begins. In order to maximize capacity, the voltage variation tolerance should be less than ¡À1\%. The current will decrease towards 20mA.

Stage 4: Charge Termination (CT) - Two methods approaches are typically used to terminate charging: 1) minimum current charge or 2) timer cutoff. 
In the minimum current approach, battery charge is terminated when the current diminishes below 20mA, the minimum current threshold, during the CV stage. In the timer cutoff approach, for example, 2-hour timer starts when the CV stage is invoked. The charge is terminated after 2-hour during the CV stage.

In summary, the charge profile is the desired charging voltage-current pair at a given time in charging duration.

\subsection{Feedback System in ADLC}\label{}
At the ADLC receiver, the power monitor can track the battery voltage and current. Based on the battery charge profile and the PV-panel output power, the desired laser power can be figured out, which will be discussed below. The power monitor can send the desired laser power to the power controller via the feedback channel.

The feedback information channel can be established using various wireless communication technologies, e.g., WiFi, Bluetooth, infra-communication, etc..

At the ADLC transmitter, the feedback information is handled by the power controller. Power controller adjusts the power supply to provide different stimulation power on the gain medium to generate different resonating beam power. Thus, this adaptation procedure leads to the corresponding laser power control.

\subsection{Electricity-to-Laser Power Conversion}\label{}
Based on laser physics, only when stimulation current is over a certain threshold, laser can be generated \cite{Threshold-Emperature}. We denote $I_{t}$ as the stimulation current applied on the gain medium, which is provided by the power supply. At the ADLC receiver, we denote $P_{laser}$ as the laser beam power.

It is well-known in the laser diode physics that the $P_{laser}$ depends on $I_{t}$, which can be depicted as \cite{photonic}:
\begin{equation}\label{powervscurrent}
P_{laser}= \frac{h\upsilon}{q}(I_{t}-I_{th}),
\end{equation}
where $q$ is the elementary charge constant, $h$ is the Planck constant, $\upsilon$ is the constant laser frequency and $I_{th}$ is the constant current threshold determined by the gain medium.

\subsection{Laser-to-Electricity Power Conversion}\label{}
\begin{figure}
	\centering
	\includegraphics[width=2.4in]{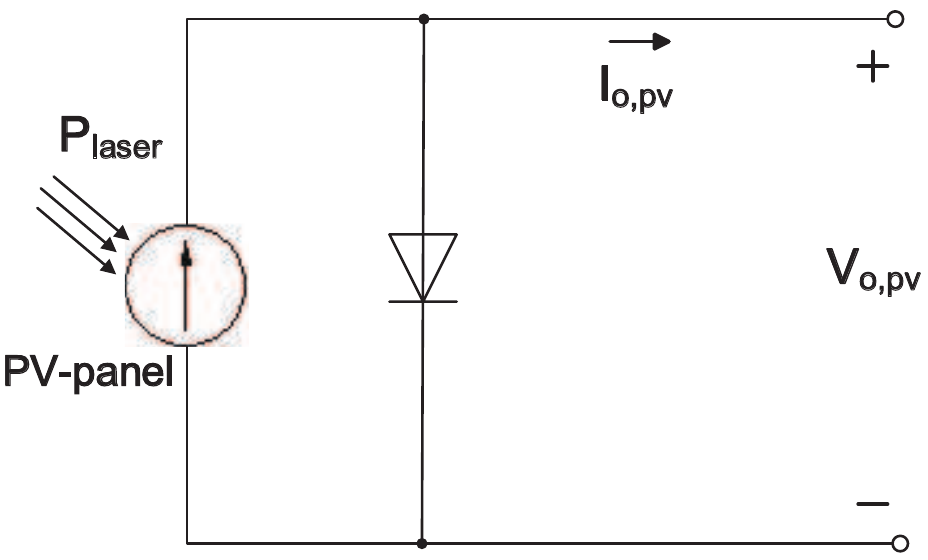}
	\caption{PV-panel Power Conversion Circuit Model}
	\label{single-diode}
\end{figure}

At the ADLC receiver, laser beam power is converted to electrical power via PV-panel \cite{Aziz2014}. Fig. \ref{single-diode} is the circuit model of semiconductor. The PV-panel output voltage $V_{o,pv}$ and current $I_{o,pv}$ can be characterized as:
\begin{equation}\label{pvconversion}
  I_{o,pv}=I_{sc} - I_{s}(e^{{V_{o,pv}}/{V_t}}-1), 
\end{equation}
where $I_{sc}$ is the PV-panel short-circuit current, and $I_s$ is the ``saturation current'', i.e., the diode leakage current density in the absence of light, which depends on the open-circuit voltage, which is just $V_{o,pv}$, as:
\begin{equation}\label{Is}
  I_s=I_{sc}e^{-V_{o,pv}/V_t}.
\end{equation}
The ``thermal voltage'' $V_t$ is defined as:
\begin{equation}\label{Vt}
  V_t=nkT/q,
\end{equation}
where $n$ is PV-panel quality factor, $k$ is Boltzmann constant, $q$ is electron charge constant, and $T$ is absolute temperature.

In summary, PV-panel converts the laser beam power $P_{laser}$ to electrical current $I_{o,pv}$ and voltage $V_{o,pv}$ at the ADLC receiver. There is no closed-form expression, however, the $I_{o,pv}$ and $V_{o,pv}$ coupled values can be determined given $P_{laser}$. Furthermore, the maximum power point (MPP) of PV-panel output is unique, which will be demonstrated in the performance evaluation later.

\subsection{DC-to-DC Conversion}\label{}
The PV-panel output current $I_{o,pv}$ and voltage $V_{o,pv}$ may not be optimal for the battery charge profile. Therefore, we need a method for converting $I_{o,pv}$ and $V_{o,pv}$ to the desired charging values with the minimum power loss.

In solar power systems, it is well-known that the DC-DC converter \cite{DC-DC converter} between PV-panel and the load can optimize power conversion efficiency in the case of dynamic solar radiation and different power load requirements. Similarly, the DC-DC converter is used in the ADLC receiver.

\begin{figure}
	\centering
	\includegraphics[width=3.45in]{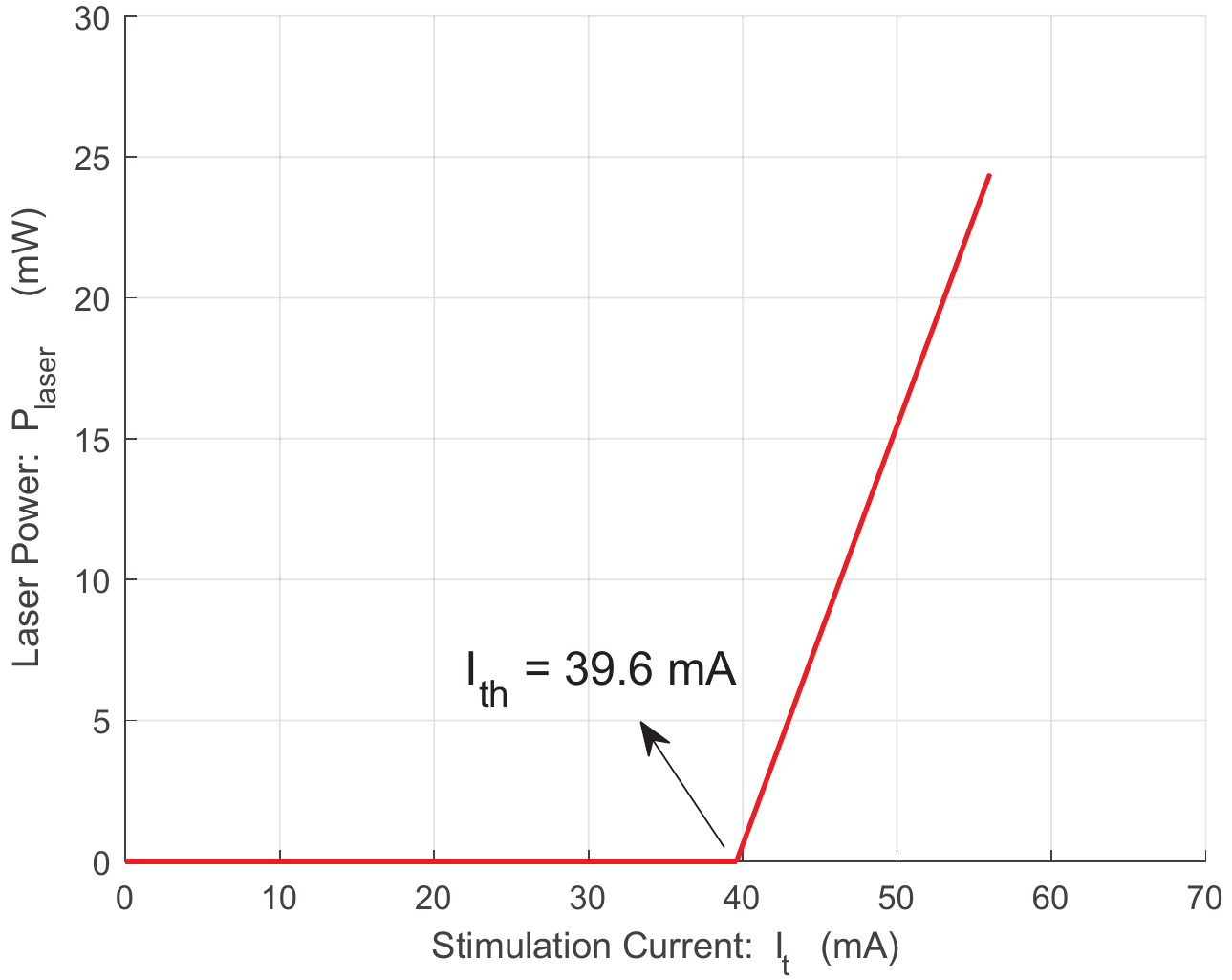}
	\caption{Laser Power vs. Stimulation Current}
	\label{semiconductor}
\end{figure}

At the ADLC receiver, the DC-DC converter is a programmable integrated circuit, which can transform the PV-panel voltage $V_{o,pv}$ and current $I_{o,pv}$ to the desired battery charging voltage $V_{o,dc}$ and current $I_{o,dc}$. The DC-DC converter output current and voltage are controllable, which provides the flexibility to optimize battery charging.

\subsection{ADLC Operation}\label{}
At the ADLC receiver, the desired battery charging power should be the product of voltage and current in the battery charge profile, which is tracked by the power monitor. The DC-DC converter input and output power should equal to this desired battery charging power. It is also the PV-panel output power. The desired laser power can be figured out according to corresponding values of voltage and current of certain PV-panel MPP. The power monitor will send the desired laser power to the ADLC transmitter via feedback channel. The power controller will take this information to adjust power supply to generate the desired laser power.


\section{Performance Evaluation}\label{}
Table \ref{paramaters} lists the simulation parameters in our performance evaluation, which is based on MATLAB and Simulink.

\subsection{Electricity-to-Laser Power Conversion}\label{}
We consider here an InGaAsP/InP double heterostructure laser diode \cite{photonic}. Based on Eq. \eqref{powervscurrent}, the laser power generated by stimulation current is shown in Fig. \ref{semiconductor}, which illustrates the linear relationship between $P_{laser}$ and $I_{t}$ when $I_{t}>I_{th}$.

\begin{figure}
	\centering
	\includegraphics[width=3.5in]{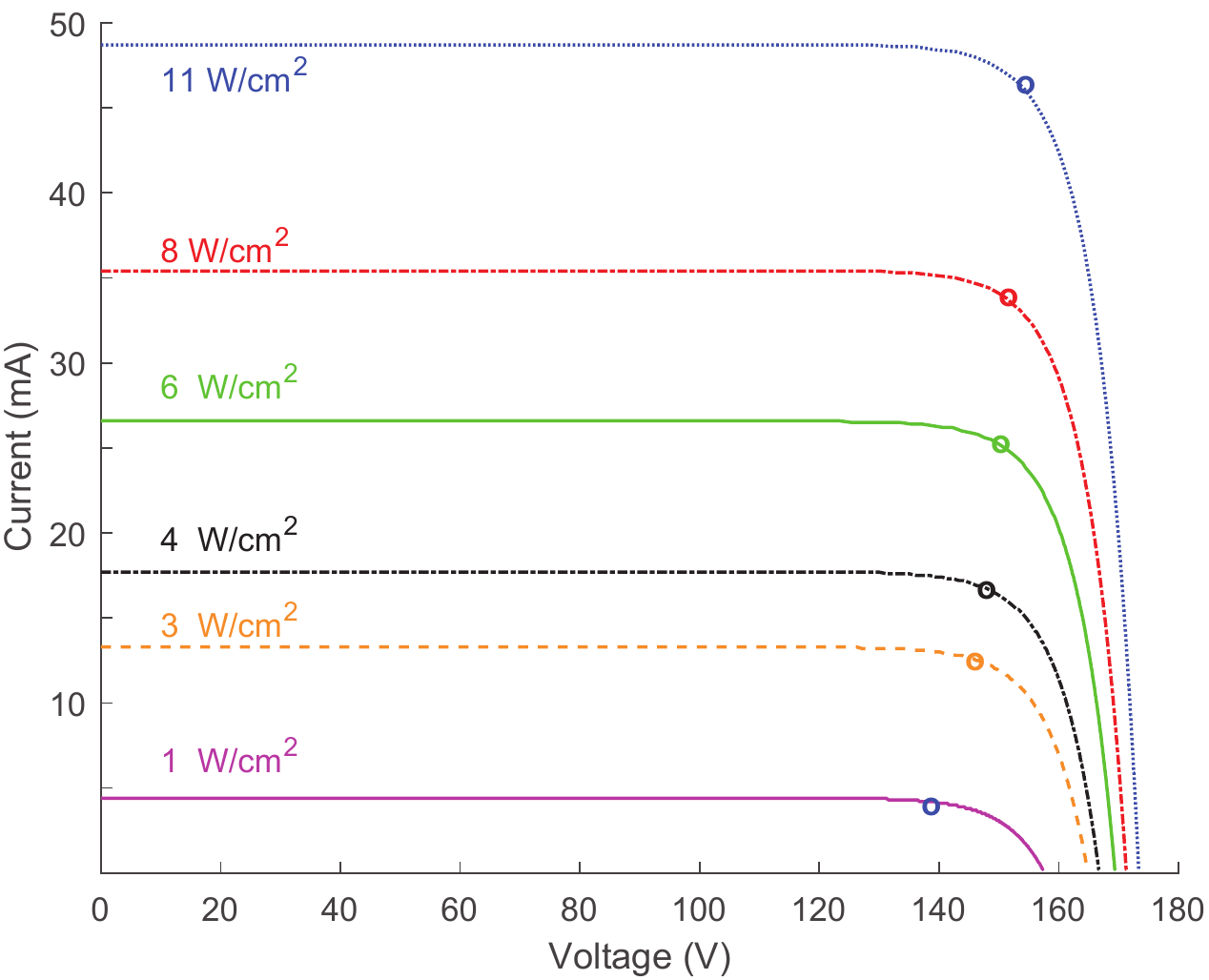}
	\caption{PV-panel Output Current vs. Voltage}
	\label{Iirradiance}
\end{figure}

\subsection{Laser-to-Electricity Power Conversion}\label{}
PV-panel converts laser to electricity. Based on Eq. \eqref{pvconversion} to \eqref{Vt}, Fig.~\ref{Iirradiance} depicts the relationship between the PV-panel output current $I_{o,pv}$ and voltage $V_{o,pv}$. Given $V_{o,pv}$, the higher the input laser power, the higher the output current $I_{o,pv}$.
\begin{table}[bp]
\centering
\caption{Parameter Setting}
\begin{tabular}{C{3.2cm} C{0.4cm} C{3.2cm}}
\hline
 Parameter & & Value  \\
\hline
\bfseries{Absolute temperature}                       & {T}           & $298 K$ \\
\bfseries{Planck constant}                            & {h}           & $6.62606957\times10^{-34} J\cdot s$ \\
\bfseries{Laser frequency}                            & {v}           & $3.59\times10^{14} Hz$ \\
\bfseries{Electron charge}                            & {q}           & $1.6\times10^{-19} C$ \\
\bfseries{Current threshold}                          & {I$_{th}$}    & $39.6 mA$ \\
\bfseries{Boltzmann's constant}                       & {k}           & $1.38064852\times10^{-23} J/K $ \\
\bfseries{Short-circuit current}                      & {I$_{sc}$}    & $128 mA$ \\
\bfseries{Open-circuit voltage}                       & {V$_{o,pv}$}  & $5.99 V$ \\
\bfseries{Irradiance used for measurements}           & {I$_{r0}$}    & $28.9 W/cm^2$ \\
\bfseries{Quality factor}                             & {n}           & $8.5$\\
\bfseries{Number of series cells}                     & {N}           & $30$ \\
\hline
\label{paramaters}
\end{tabular}
\end{table}

\subsection{DC-DC Conversion}\label{}

Fig.~\ref{Pirradiance} shows the relationship between the PV-panel output power $P_{o,pv}=I_{o,pv}V_{o,pv}$ and voltage $V_{o,pv}$, given a certain laser power density. The MPP is the peak value point, which is marked by a circle. To obtain the optimal PV-panel conversion efficiency, the PV-panel output voltage and current corresponding to the MPP should be found given laser power. On the other hand, if the desired PV-panel output power is given, based on MPP, the unique laser power can be determined.

Since DC-DC conversion is ideal, the PV-panel output power is same as battery charging power. Therefore, given the optimal battery charging power, i.e., the optimal PV-panel output power, the optimal laser power, which is the corresponding MPP, can be tracked out. This procedure of tracking battery charging profile can be called the maximum power point tracking (MPPT).

\subsection{ADLC Operation}\label{}
\begin{figure}
	\centering
	\includegraphics[width=3.5in]{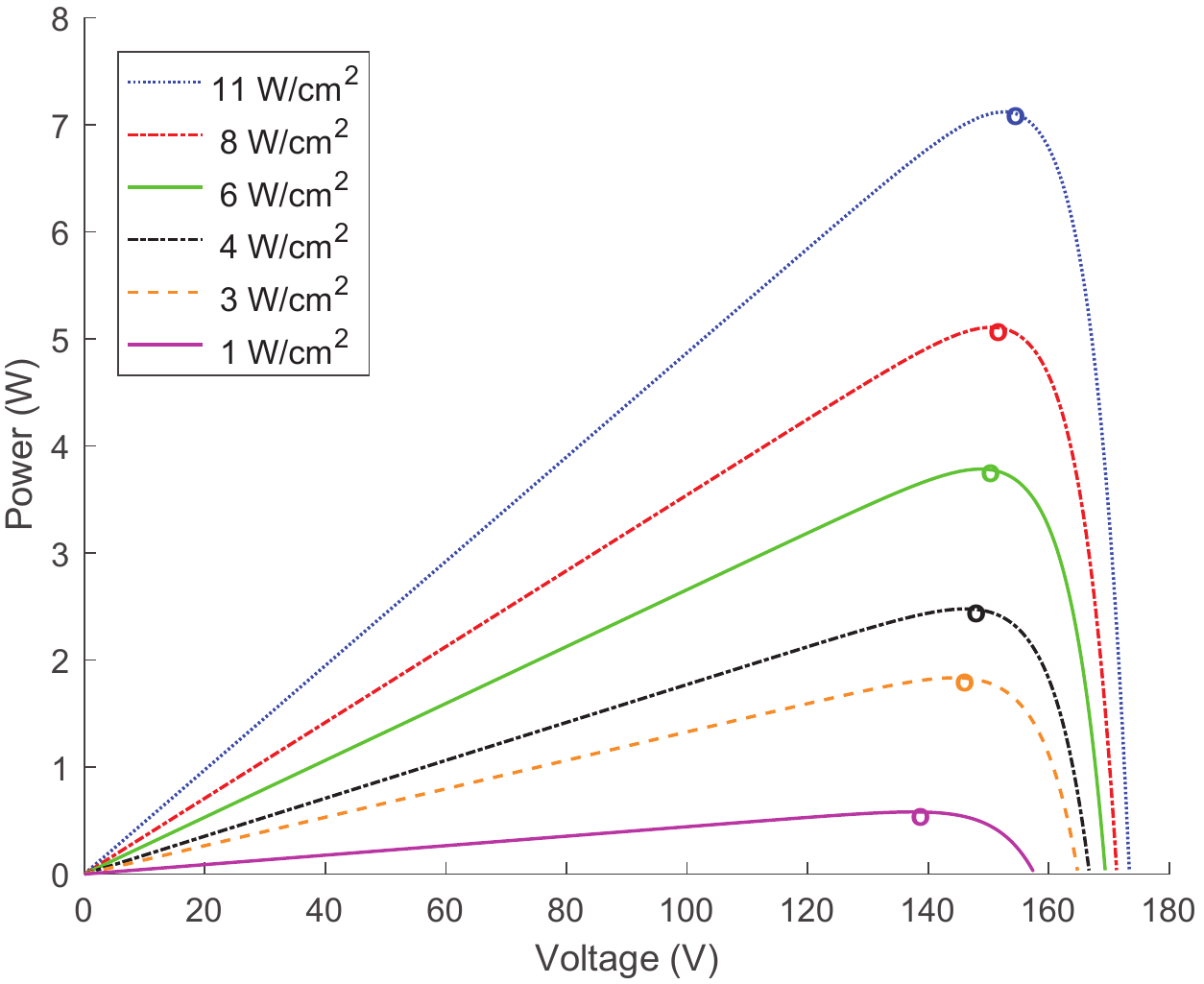}
	\caption{PV-panel Output Power vs. Voltage}
	\label{Pirradiance}
\end{figure}

An example is presented here to demonstrate the ADLC operation. We assume that the battery is charged at the end of the CC stage, according to Fig. \ref{li-ionchargezero}. The desired current and voltage are 600mA and 4.2V, respectively. So the desired DC-DC output power, i.e., PV-panel output power, should be 600mA$\times$4.2V$=$2.52W. After searching Fig. \ref{Pirradiance}, the desired laser irradiation can be figured out as about 4W/cm$^2$, where the corresponding PV-panel output current and voltage are 17mA and 148V, which can be obtained from Fig. \ref{Iirradiance}, respectively. The desired laser power density 4W/cm$^2$ is provided to the power supply through feedback. Then, the desired laser power can be generated according to Fig. \ref{semiconductor}.

\subsection{Performance Comparison}\label{}

\begin{figure}
	\centering
	\includegraphics[width=3.5in]{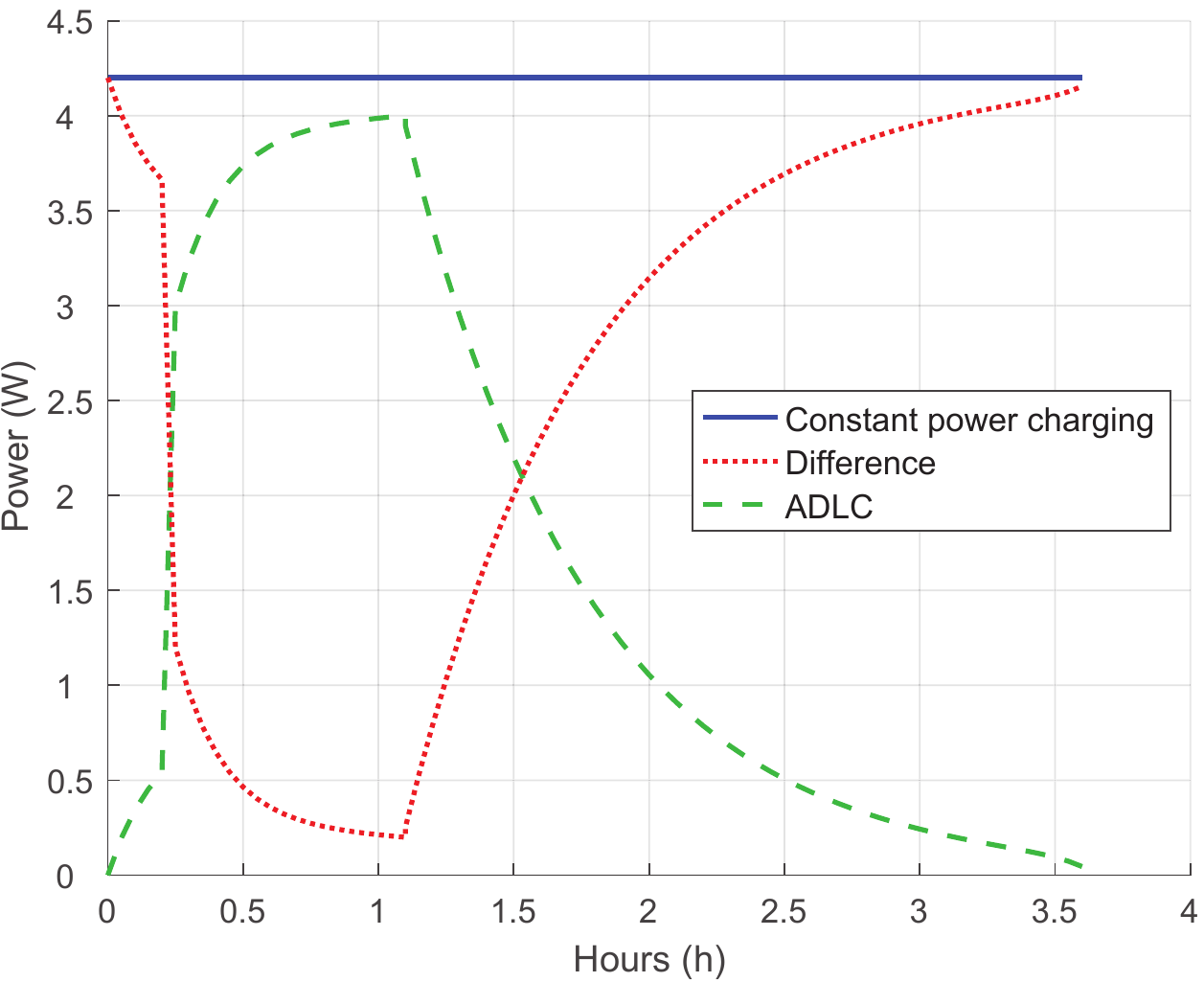}
	\caption{Battery Charging Power}
	\label{efficiency}
\end{figure}

In most traditional wireless charging systems including the DLC system, batteries can only be charged at a fixed power level. Based on the Li-ion battery charge profile in Fig. \ref{li-ionchargezero}, the minimum power 4.2W (constant voltage 4.2V and constant current 1A) is needed for Li-ion battery charging. The solid-line in Fig. \ref{efficiency} shows the constant 4.2W charging power.

Battery charging power in the ADLC system is depicted by the dashed-line in Fig. \ref{efficiency}, relying on the perfect battery profile tracking as in Fig. \ref{li-ionchargezero}. The gap between the constant power charging and ADLC is illustrated by a dotted-line, which is the saved power by ADLC.

The integral of power over time yields energy. After computation, we find that the constat power charging consumes 15.12wh and ADLC takes 5.98wh. The difference is 9.14wh, which means at least 60.4\% of energy can be saved by ADLC over the traditional fixed-power charging system.

\section{Conclusion}

In this paper, an adaptive distributed laser charging system is introduced to transmit wireless power for mobile devices. It can automatically adjust the transmitting laser power based on feedback information, according to dynamic battery charge profile. In this system, a direct current to direct current converter is adopted to deliver the optimal voltage and current for battery charging. The optimal laser power utilization and battery charging performance was achieved. Performance evaluation illustrates that at least 60.4\% of energy can be saved by using our proposed system, comparing to the traditional constant power charging system.

There are still some open issues to be studied in the future. For example, the impact of resonating laser beam propagation loss should be further investigated. The temperature effects on the system could be evaluated as well.


\end{document}